# An artificial atom locked to natural atoms

N. Akopian[1*],   R. Trotta[2],   E. Zallo[2],   S. Kumar[2],   P. Atkinson[2†],   A. Rastelli[2], O. G. Schmidt[2] & V. Zwiller[1]

[1]*Kavli Institute of Nanoscience Delft, Delft University of Technology, 2628CJ Delft, The Netherlands.*

[2]*Institute for Integrative Nanosciences, IFW Dresden, 01069 Dresden, Germany.*

8 February 2013

**Single-photon sources that emit photons at the same energy play a key role in the emerging concepts of quantum information[1], such as entanglement swapping[2], quantum teleportation[3,4] and quantum networks[5]. They can be realized in a variety of systems[6,7,8,9,10,11,12], where semiconductor quantum dots, or 'artificial atoms'[13], are arguably among the most attractive. However, unlike 'natural atoms', no two artificial atoms are alike. This peculiarity is a serious hurdle for quantum information applications that require photonic quantum states with identical energies. Here we demonstrate a single artificial atom that generates photons with an absolute energy that is locked to an optical transition in a natural atom. Furthermore, we show that our system is robust and immune to drifts and fluctuations in the environment of the emitter. Our demonstration is crucial for**

---

\* To whom correspondence should be addressed. E-mail: n.akopian@tudelft.nl
† Present address: *Institut des Nanosciences de Paris, Université de Pierre et Marie Curie, CNRS UMR 7588, 4 Place Jussieu, Boîte Courrier 840, 75005 Paris, France.*

**realization of a large number of universally-indistinguishable solid-state systems at arbitrary remote locations, where frequency-locked artificial atoms might become fundamental ingredients.**

Quantum dots are robust emitters of on-demand single[14] and entangled photons[15], with high repetition rates and narrow spectral linewidths. Moreover, quantum dots can be naturally integrated into larger nano-fabricated structures[16,17], compatible with modern nanoelectronics. Recently, two dissimilar quantum dots (QDs) have been tuned to the same energy relative to each other, allowing for two-photon interference of their emission[9,11]. However, scaling of this concept to more than two QDs at distant locations becomes substantially complicated and non-trivial, since all QDs in the network have to be tuned relative to each other. It would be desirable to be able to precisely tune the emission energy of each QD independently, for instance, to an absolute energy reference.

Here we demonstrate a hybrid semiconductor-atomic system where an individual artificial atom generates single photons with an absolute energy. We adopt the concept of atomic clocks[18], where the frequency of an atomic transition is used as a frequency standard for the time-ticking element of the clock. Likewise, in our work we tune the emission frequency (or, equivalently, the emission energy) of a QD in resonance with an atomic transition[19,20], as illustrated in Fig. 1. We then lock the QD emission by implementing a feedback loop that monitors and minimizes the detuning from the resonance. Our scheme can, therefore, provide a powerful solution to the mismatch in the photon energies of QDs and enable independent tuning of multiple QDs to the absolute energy of the atomic transition, immune to drifts and relative calibrations.

We first outline the relevant characteristics of the two systems that we couple: a single GaAs QD (an artificial atom) and rubidium (a natural atom). Our GaAs QDs were specially designed to generate single photons around 780 nm, close to the $^{87}$Rb $D_2$ transitions[21,22,23] (see Supplementary Fig. S1 and Methods). To increase photon extraction efficiency, the QDs were integrated in an optical micro-cavity. The QD emission wavelength (or, equivalently, the emission energy) can be fine-tuned to rubidium transitions with external magnetic or electric fields[21], or strain[23] applied to the sample. In our experiments we use the $D_2$ transitions in $^{87}$Rb at the characteristic wavelength of 780 nm. The ground state has a hyperfine splitting composed of two absorption resonances separated by ~28 µeV.

In the experimental realization of our hybrid system, the photons emitted from a single QD propagate through the rubidium vapour before detection (see Supplementary Fig. S1 and Methods). We tune the photon energy through the $D_2$ transitions of $^{87}$Rb by applying an external magnetic field to the QD. The QD energy levels undergo a Zeeman splitting and a diamagnetic shift with increasing magnetic field. As a result, the emitted photons are controllably tuned to the $D_2$ transitions of $^{87}$Rb, and are partially absorbed by the vapour when in resonance. Our artificial atom, unlike a natural atom, is strongly coupled to its environment – electric charges, trapped in the vicinity of a QD. This charge environment fluctuates in time, modifying the local electric field and leading to unwanted detuning of the emission from the resonance. To keep the emission in resonance we have implemented an active feedback loop that controls the magnetic field in order to counteract the changes in the photons' energy (see Supplementary Methods). Additionally, we emphasize that spectrally narrow optical transitions in rubidium easily allow for QD spectroscopy with sub-µeV resolution, similar to laser spectroscopy.

In the first experiment, we take advantage of our hybrid system and demonstrate high-resolution spectroscopy of a QD. In Fig. 2a we show transmission of the two branches of a Zeeman-split emission line under increasing magnetic field. The long wavelength branch is scanned through the $D_2$ transitions and its trace is represented in Fig. 2b. We clearly resolve two transmission dips due to the hyperfine structure of the $D_2$ line, indicating a narrow spectral linewidth of our QDs. The experimental data are presented together with the model (see Supplementary Methods) with only one fit parameter – the spectral linewidth of the emission. The model fits well the experiment, revealing a very narrow, nearly lifetime-limited emission linewidth of 3.8 µeV (see Fig. S2 for lifetime measurements).

In the second experiment, shown in Fig. 3, we demonstrate tuning and locking of the QD emission energy to the atomic transition. We first scan the emission of another QD through the $D_2$ transitions and resolve two transmission dips, represented in the 'Scan' part of Fig. 3d. Next, the feedback algorithm increases the magnetic field to tune the emission to the low energy $D_2$ transition, and then regulates the field to lock and stabilize the emission energy at the resonance, as shown in the 'Tune & lock' part of Fig. 3a. Transmissions of the two Zeeman-split branches of the QD emission line, reference (Fig. 3b) and signal (Fig. 3c), are measured simultaneously. The feedback algorithm monitors and minimizes the transmission ratio of the signal and reference lines (Fig. 3d), therefore relying on the intensity fluctuations that originate only from the resonance detuning (see also Fig. S3). Studying the energy distribution of photons detuned from the resonance in Fig. 3a we obtain an accurate locking, within 1/3 of the emission linewidth (for details see Fig. S4 and Fig. S5). The locking precision can be further increased by realizing, for instance, a faster and more sophisticated analogue feedback algorithm that monitors the derivative of the transmission, which is more sensitive to changes in the emission energy.

In the final part of the experiment, shown in the 'Re-lock' part of Fig. 3, we probe the charge environment in the vicinity of the QD. First we increase the excitation intensity by more than 3 orders of magnitude for 5 s, indicated by the red line. This very intense, non-resonant illumination redistributes electrons and holes trapped by the impurities close to the QD, and therefore creates a new charge environment, that modifies the local electric field. The energy levels in the QD are thus changed, resulting in a spectral jump and in the immediate detuning of the emission from the resonance, as represented by an abrupt rise of signal transmission in Fig. 3d. The feedback loop then tunes and re-locks the emission, but now at higher magnetic field, to compensate for the change in the local electric field. The difference in the magnetic field corresponds to a 34 μeV shift in the emission energy that we deduce directly from the experimental data in Fig. 3a. This energy shift corresponds to the net change of only a few single charges in the vicinity of the QD[24,25], demonstrating the sensitivity of our technique to detect elementary charges. The spikes in the signal transmission, marked by arrows in Fig. 3d, represent another evidence of spontaneous fluctuations of single charges, that we entirely suppress with our feedback loop (for details see Fig. S6). We therefore demonstrate that frequency-locking of a QD provides a robust solution for energy stabilization, immune to fluctuations and drifts caused by the environment, such as slow spectral wandering.

Our results, especially the strong sensitivity to detuning from resonance, immediately suggest several experiments. For instance, our method can enable detection of single charges and study of the dynamics of such charge environment. Furthermore, with the implementation of faster analogue feedback, our system could counteract the fluctuations of the environment considerably stronger, significantly reducing the harmful spectral wandering of QD emission[21]. Finally, a rapid and on-chip tuning of QD energy levels can be realized exploiting the Stark effect[21] or applying strain[23,26] to the sample.

We have demonstrated a powerful solution to one of the major hindrances in application of QDs to quantum networks, teleportation and quantum repeaters[27] – single QDs that emit photons with an absolute energy – by precise tuning and frequency-locking the photon energy to an atomic transition. More generally, our hybrid semiconductor-atomic system can now open exciting new avenues, where the advantages of both constituents are combined together: the scalability and functionality of nanostructure devices based on single QDs and the uniformity of atomic systems.

**References**

1.  Bouwmeester, D., Ekert, A. K. & Zeilinger, A. *The Physics of Quantum Information* (Springer, 2000).

2. de Riedmatten, H. *et al.* Long-distance entanglement swapping with photons from separated sources. *Phys. Rev. A* **71**, 050302 (2005).

3. Bouwmeester, D. *et al.* Experimental quantum teleportation. *Nature* **390**, 575-579 (1997).

4. Furusawa, A. *et al.* Unconditional quantum teleportation. *Science.* **282**, 706-709 (1998).

5. Kimble, H. J. The quantum internet. *Nature* **453**, 1023-1030 (2008).

6. Beugnon, J. *et al.* Quantum interference between two single photons emitted by independently trapped atoms. *Nature* **440**, 779-782 (2006).

7. Maunz, P. *et al.* Quantum interference of photon pairs from two remote trapped atomic ions. *Nature Phys*. **3**, 538-541 (2007).

8. Sanaka, K., Pawlis, A., Ladd, T. D., Lischka, K. & Yamamoto, Y. Indistinguishable photons from independent semiconductor nanostructures. *Phys. Rev. Lett.* **103**, 053601 (2009).

9. Flagg, E. B. *et al.* Interference of single photons from two separate semiconductor quantum dots. *Phys. Rev. Lett.* **104**, 137401 (2010).

10. Lettow, R. *et al.* Quantum interference of tunably indistinguishable photons from remote organic molecules. *Phys. Rev. Lett.* **104**, 123605 (2010).

11. Patel, R. B. *et al.* Two-photon interference of the emission from electrically tunable remote quantum dots. *Nature Photon.* **4**, 632-635 (2010).

12. Bernien, H. *et al.* Two-photon quantum interference from separate nitrogen vacancy centers in diamond. *Phys. Rev. Lett.* **108**, 043604 (2012).

13. Kastner, M. A. Artificial atoms. *Physics Today* **46**, 24-31 (1993).

14. Michler, P. *et al.* A quantum dot single-photon turnstile device. *Science* **290**,

2282–2285 (2000).

15. Akopian, N. *et al.* Entangled photon pairs from semiconductor quantum dots. *Phys. Rev. Lett.* **96**, 130501 (2006).

16. Krenner, H.J. *et al.* Direct observation of controlled coupling in an individual quantum dot molecule. *Phys. Rev. Lett.* **94**, 057402 (2005).

17. Claudon, J. *et al.* A highly efficient single-photon source based on a quantum dot in a photonic nanowire. *Nature Photon.* **4**, 174-177 (2010).

18. Essen, L. & Parry, J. V. L. An atomic standard of frequency and time interval: a cæsium resonator. *Nature* **176**, 280-282 (1955).

19. Liu, C., Dutton, Z., Behroozi, C.H. & Hau, L.V. Observation of coherent optical information storage in an atomic medium using halted light pulses. *Nature* **409**, 490-493 (2001).

20. Huber, B. *et al.* GHz Rabi flopping to Rydberg states in hot atomic vapor cells. *Phys. Rev. Lett.* **107**, 243001 (2011).

21. Akopian, N. *et al.* Tuning single GaAs quantum dots in resonance with a rubidium vapor. *Appl. Phys. Lett.* **97**, 082103 (2010).

22. Akopian, N., Wang, L., Rastelli, A., Schmidt, O. G. & Zwiller, V. Hybrid semiconductor-atomic interface: slowing down single photons from a quantum dot. *Nature Photon.* **5**, 230-233 (2011).

23. Kumar, S. *et al.* Strain-induced tuning of the emission wavelength of high quality GaAs/AlGaAs quantum dots in the spectral range of the $^{87}$Rb $D_2$ lines. *Appl. Phys. Lett.* **99**, 161118 (2011).

24. Vamivakas, A. N. *et al.* Nanoscale optical electrometer. *Phys. Rev. Lett.* **107**, 166802 (2011).

25. Houel, J. *et al.* Probing single charge fluctuations in a semiconductor with laser

spectroscopy on a quantum dot. *Phys. Rev. Lett.* **108**, 107401 (2012).

26.  Trotta, R. *et al.* Nanomembrane Quantum-Light-Emitting Diodes Integrated onto Piezoelectric Actuators. *Adv. Mater.* **24**, 2668 (2012).

27.  Briegel, H.-J., Dür, W., Cirac, J. I. & Zoller, P. Quantum repeaters: the role of imperfect local operations in quantum communication. *Phys. Rev. Lett.* **81**, 5932-5935 (1998).

**Acknowledgements** This work was supported by the Dutch Organization for Fundamental Research on Matter (FOM), The Netherlands Organization for Scientific Research (NWO Veni/Vidi), the SOLID program, the DFG (FOR 730) and the BMBF (QK_QuaHL-Rep, 01BQ1032).

**Author Contributions** The experiments were conceived and designed by N.A. and V.Z. and carried out by N.A. The data were analysed and modelled by N.A. The sample was developed by R.T., E.Z., S.K., P.A., A.R. and O.G.S. The manuscript was written by N.A. with input from all co-authors.

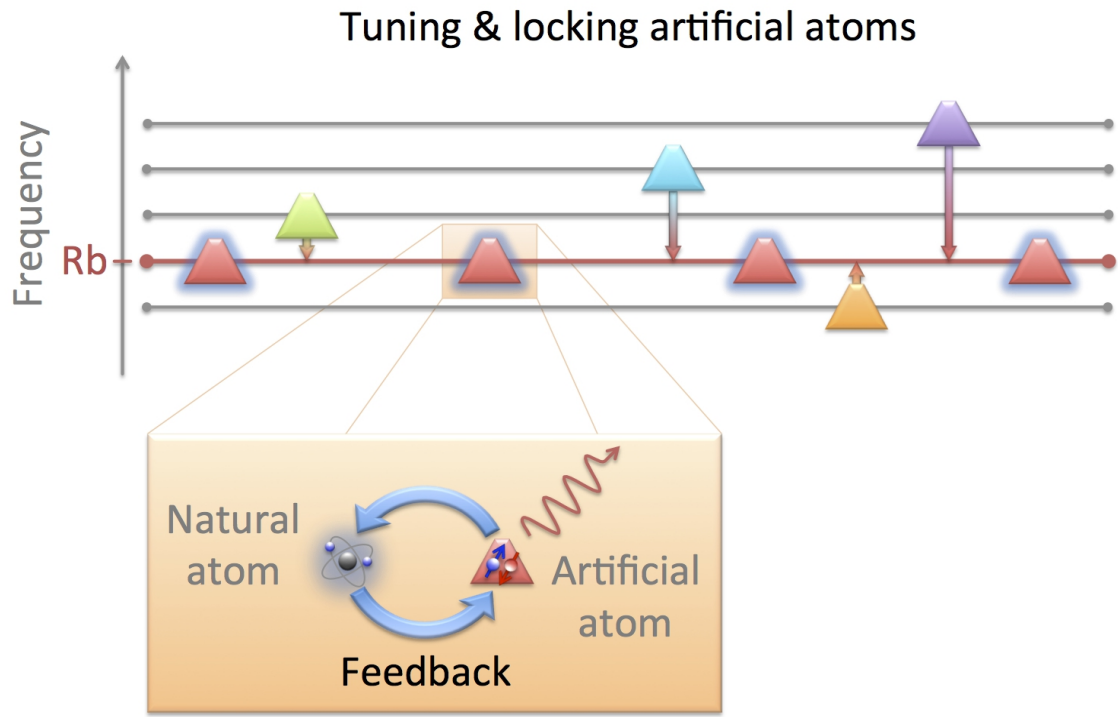

**Fig. 1 legend | Locking an artificial atom to natural atoms.** Artificial atoms, represented as pyramids, are all different. They emit photons at different frequencies, except those that are tuned and locked to the frequency of an optical transition in a natural atom – $^{87}$Rb. An active feedback algorithm ensures that an artificial atom is locked to a natural atom at all times, and therefore generates single photons with an absolute energy, equal to the optical transition of $^{87}$Rb.

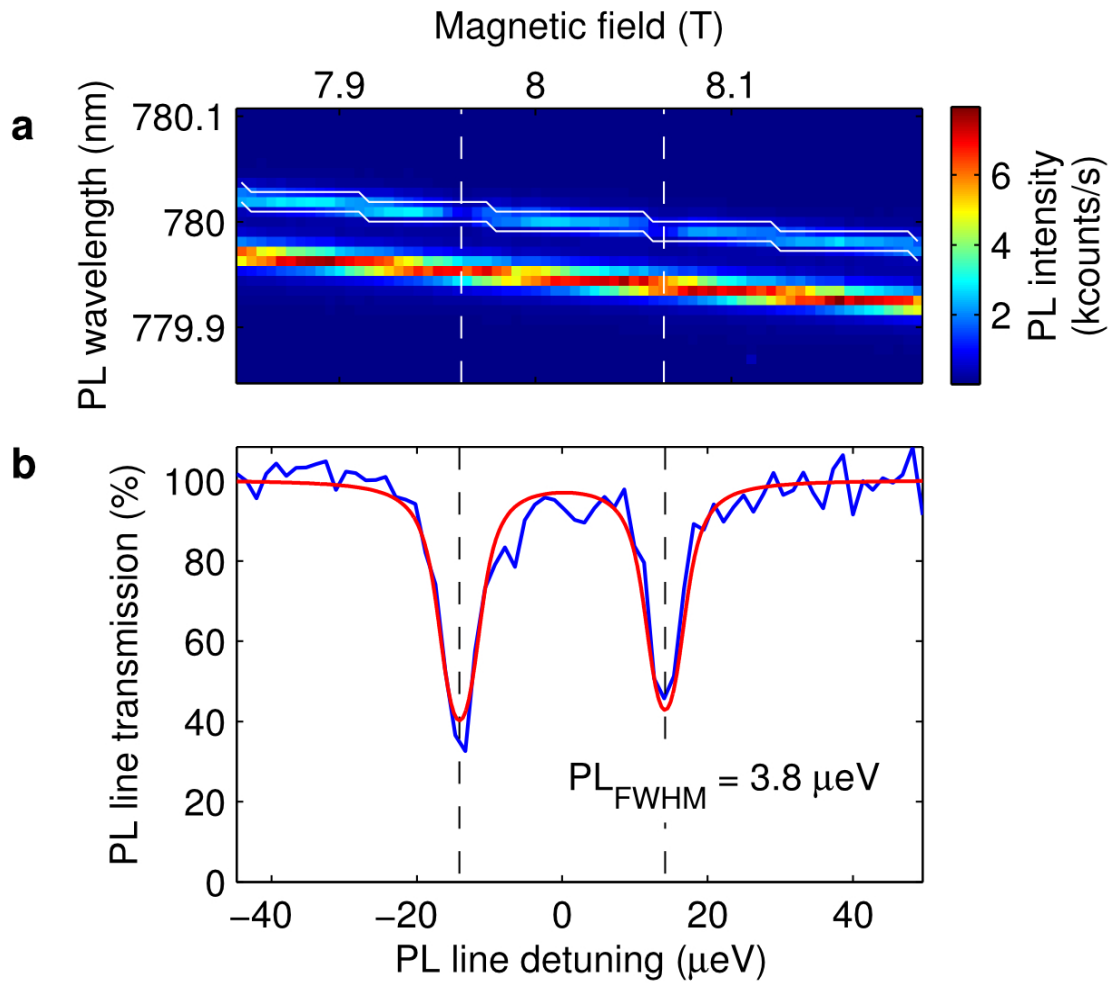

**Fig. 2 legend | High-resolution spectroscopy of a QD. a,** Photoluminescence (PL) spectra of a QD exciton under increasing magnetic field. The QD was excited with a 532 nm continuous-wave laser at 100 nW. A cell with a rubidium vapour at 81 °C was placed in front of the detector. Dashed lines correspond to the [87]Rb $D_2$ transitions. The long wavelength branch of the Zeeman-split emission, marked by solid white lines, is tuned through the $D_2$ transitions and is partially absorbed by the vapour when in resonance. **b,** Transmission of the PL line through the rubidium vapour: experimental data (blue) and fit of our model (red). The experimental data is obtained by tracing the long wavelength PL line in (**a**) (see Fig. S3 for details). The abscissa units are converted from Tesla to µeV (see Supplementary Methods). The data show two well-resolved dips

that correspond to the $D_2$ hyperfine structure, indicated by dashed lines. The fit reveals very narrow emission linewidth with full width at half maximum (FWHM) of 3.8 ± 0.3 µeV.

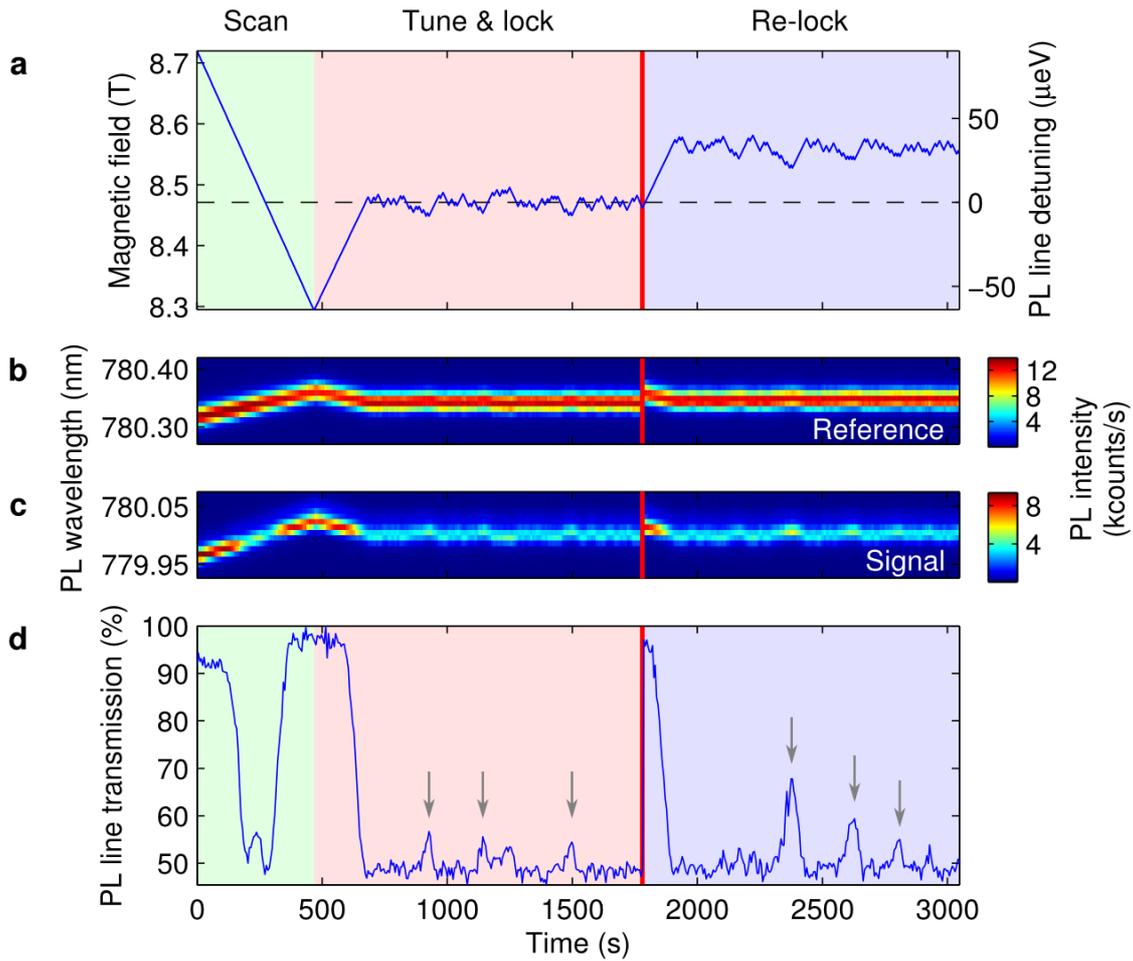

**Fig. 3 legend | Locking of the QD emission energy to an atomic transition. a,** Modulation of the magnetic field applied to the sample, expressed also as the detuning of the PL line from a rubidium transition, indicated by the dashed line. **b-c,** Transmission of two PL lines: the reference and the signal, non-resonant and resonant with the $D_2$ transitions, respectively. The signal emission is partially absorbed by the rubidium vapour when in resonance. The QD was excited with a 532 nm continuous-wave laser at 60 nW. **d,** Transmission ratio of the normalized signal and the reference lines. The experiment is divided into three parts. Scan – the magnetic field is monotonically decreased to scan the QD emission through the $D_2$ transitions, represented by two dips in (**d**). Tune & lock – the feedback algorithm controls the magnetic field that is first monotonically increased to tune the emission energy into

resonance with the low energy $D_2$ transition, and is then regulated to lock and stabilize the emission energy. Re-lock – tuning and locking the emission energy after intense excitation at 0.4 mW for 5 s, indicated by the solid red line. The arrows mark the detection of spontaneous charge fluctuations in the vicinity of the QD. The rubidium vapour temperature was 122 °C.

# Supplementary Information for

# An artificial atom locked to natural atoms

N. Akopian[1], R. Trotta[2], E. Zallo[2], S. Kumar[2], P. Atkinson[2], A. Rastelli[2], O. G. Schmidt[2] & V. Zwiller[1]

[1]*Kavli Institute of Nanoscience Delft, Delft University of Technology, 2628CJ Delft, The Netherlands.*

[2]*Institute for Integrative Nanosciences, IFW Dresden, 01069 Dresden, Germany.*

**Supplementary Methods**

**Sample.** The QDs that we study were GaAs inclusions in an AlGaAs matrix, fabricated by solid source molecular beam epitaxy[23,S1]. We created a template of self-assembled nano-holes by *in situ* droplet etching on a 99 nm thick intrinsic GaAs layer. After growing a 7 nm thick $Al_{0.44}Ga_{0.56}As$ bottom barrier layer, the sample surface still showed nano-holes which were then filled with GaAs. A 2 min growth interruption promoted a net migration of the GaAs towards the bottom of the holes, giving rise to the QDs used here. Finally, the QDs were capped by a 112 nm $Al_{0.33}Ga_{0.67}As$ and 20 nm $Al_{0.44}Ga_{0.56}As$, followed by a 19 nm thick GaAs cap layer. The nominal thickness of the GaAs layer used to create the QDs was optimized to 3 nm so that the QDs photoluminescence (PL) was close to 780 nm[21,22]. The density of the QDs in the sample was low, such that an individual QD could be addressed easily. To increase photon extraction efficiency, the QDs were integrated in a metal-semiconductor-dielectric planar optical micro-cavity[26].

**Vapour cell.** We used isotopically pure $^{87}$Rb in a 75 mm long quartz cell with anti-reflection coating.

**Experimental setup.** Micro-PL studies were performed at 4.2 K in a He bath cryostat with a 9 T superconducting magnet. The QD was excited with 532 nm continuous wave or 743 nm ps pulsed lasers focused to a spot size of 1 µm using a microscope objective with numerical aperture 0.85. The PL signal was collected by the same objective and was sent through a vapour cell to a spectrometer, which dispersed the PL onto a silicon array detector or a streak camera, enabling 30 µeV spectral and 140 ps temporal resolution respectively.

**Units conversion.** We convert measurement units from Tesla to µeV by comparing the difference in the magnetic field to the corresponding energy shift of the QD emission line that we scan through the rubidium $D_2$ transitions.

**Model.** In this work we use the model that we developed recently[22]. We first calculate a frequency dependent transmission through the $D_2$ transitions, taking into account the Doppler broadening of rubidium atoms. We model a QD emission line as a Lorentzian detuned from the $D_2$ transitions. We then take a product of the rubidium transmission and the QD emission, and integrate over the frequency. The result is the transmission of the QD emission through the rubidium vapour that depends on the QD detuning. Finally, we repeat the whole procedure for QD detuning values in the range of the experiment. We fit the resulting curve to the experimental data, as shown in Fig. 2b. The FWHM of the Lorentzian, corresponding to the emission linewidth, is the only free parameter in our model.

**Feedback algorithm.** The computer controlled feedback algorithm minimizes the transmission ratio, shown in Fig. 3d, by sweeping the magnetic field applied to the sample. The feedback algorithm only changes the direction of sweep. The decision whether to change the direction is taken by comparing the transmission ratio with its previous value. If the ratio decreased – the sweep direction is correct and is not changed. If the ratio increased – the sweep direction is inverted. The size of the sweep step is kept constant during the whole process.

**Supplementary Figures**

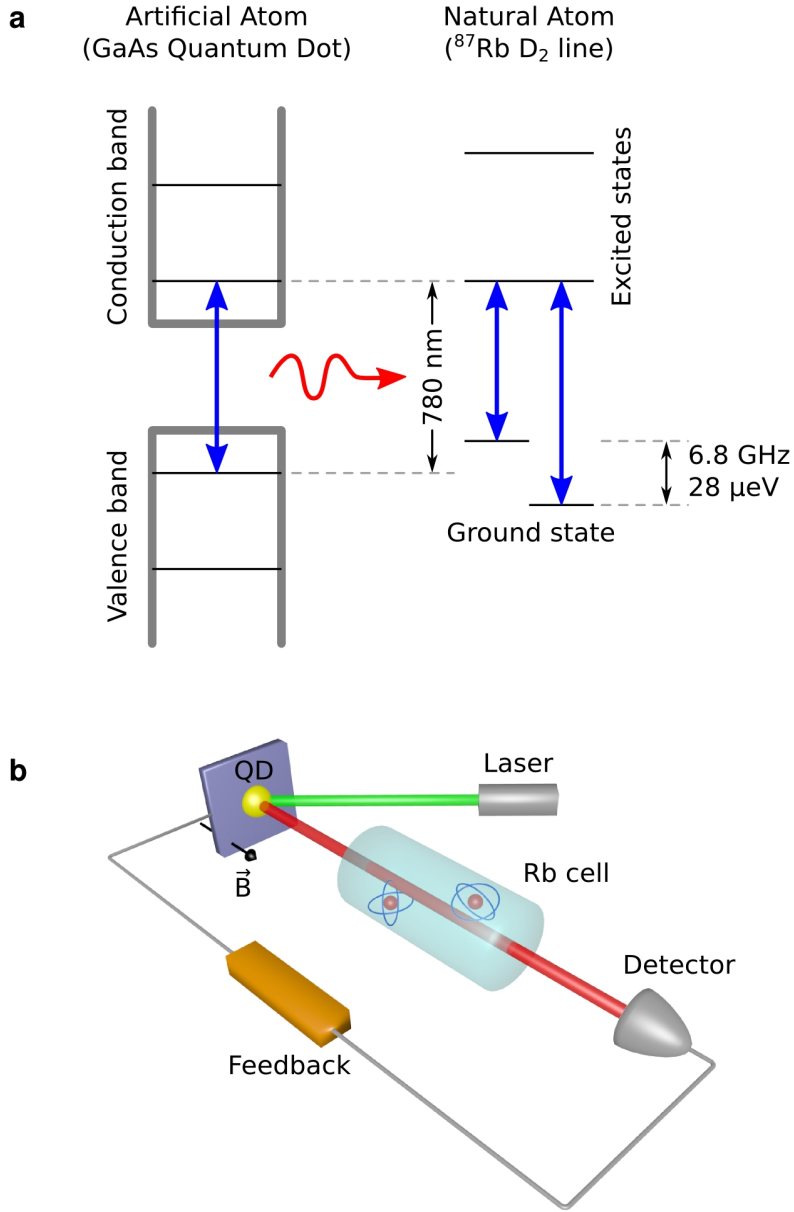

**Fig. S1 | Energy level diagrams and schematics of the experiment. a**, Energy level diagrams in QD and rubidium. The blue arrows represent relevant optical transitions. A GaAs QD is designed to emit single photons around 780 nm, close to the $^{87}$Rb D$_2$ transitions. The hyperfine structure of the $^{87}$Rb D$_2$ ground state is 28 µeV. **b**, Schematics

of the experiment. An optically excited QD emits single photons that propagate through the cell with a warm Rb vapour placed in front of the detector. The feedback algorithm based on the detection rate controls an external magnetic field applied to the sample that tunes the energy levels in a QD.

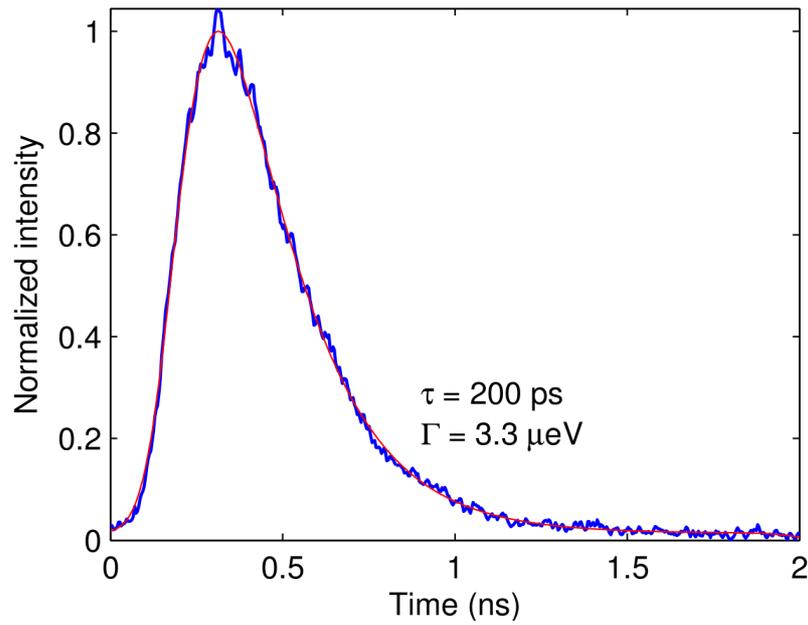

**Fig. S2 | Lifetime of an exciton.** Lifetime measurement of an exciton in a single QD within the tuning range of the $^{87}$Rb D$_2$ transitions: experimental data (blue) and fit (red). The QD was excited with a ps pulsed laser at 743 nm. The fit accounts for the temporal response of our system and has single exponential rise and decay times as parameters. The resulting decay time is $\tau = 200 \pm 3$ ps, and the corresponding natural (lifetime-limited) linewidth of the emission is $\Gamma = 3.3 \pm 0.1$ µeV.

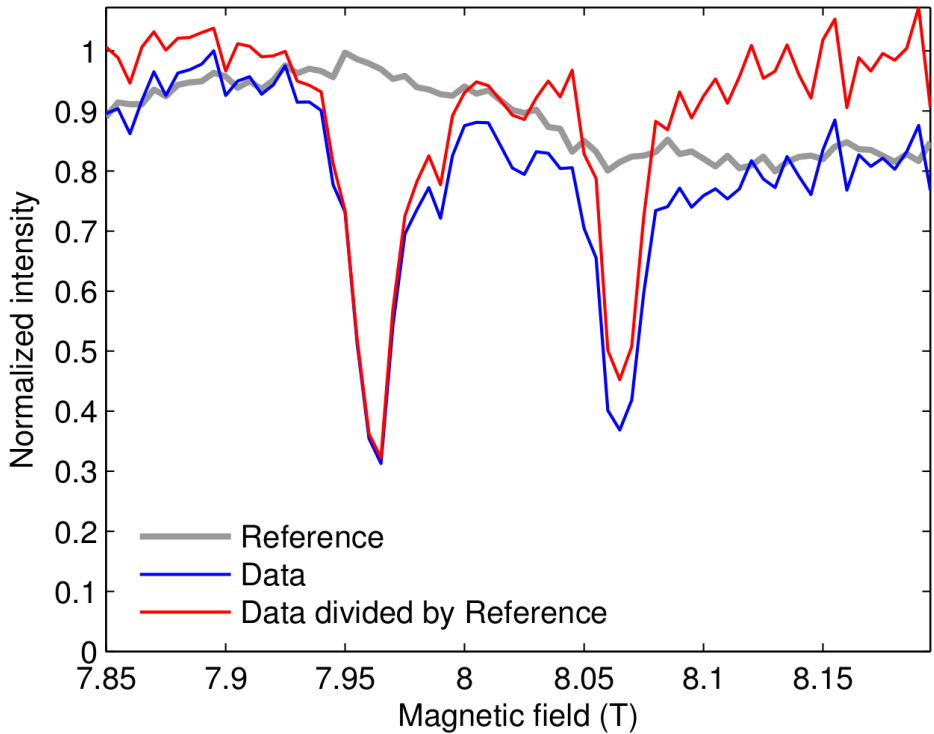

**Fig. S3 | Transmission through a vapour cell.** Normalized integrated PL intensity of a QD exciton under increasing magnetic field. Blue curve (Data) – tracing the long wavelength PL line in Fig. 2a and integrating between solid white lines. A cell with a rubidium vapour at 81 °C was placed in front of the detector. The two well-resolved dips correspond to the $D_2$ hyperfine structure. Grey curve (Reference) – the measurement of the same PL line without the rubidium cell. The PL intensity changes with the magnetic field due to relative misalignment of the sample and the objective. These intensity fluctuations are not related to the absorption by rubidium vapour. We therefore correct for them before fitting with our model. Red curve (Data divided by Reference) – the corrected data. This curve is plotted in Fig. 2b and is used for the fitting.

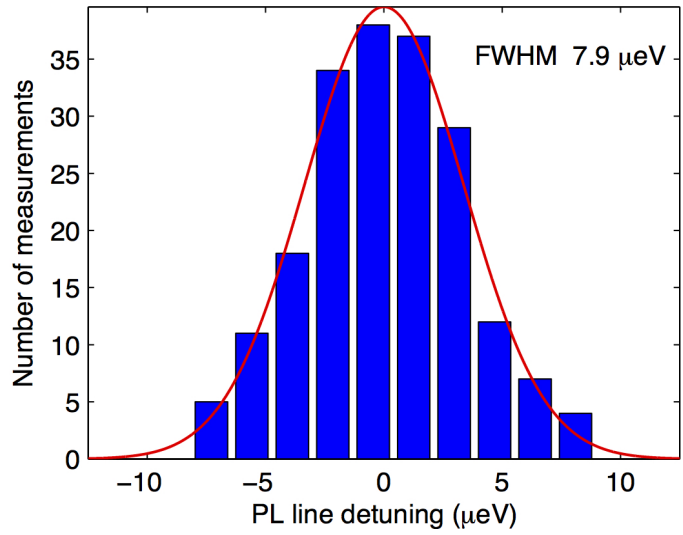

**Fig. S4 | Locking precision.** Locking precision represented as the energy distribution of photons detuned from the resonance with the atomic transition. The histogram is built on the stabilized emission energy data from Tune & lock part of Fig. 3a, from the time interval [684, 1781] s. Gaussian fit reveals the locking precision of 7.9 ± 0.4 µeV, which is 1/3 of the emission linewidth, shown in Fig. S5.

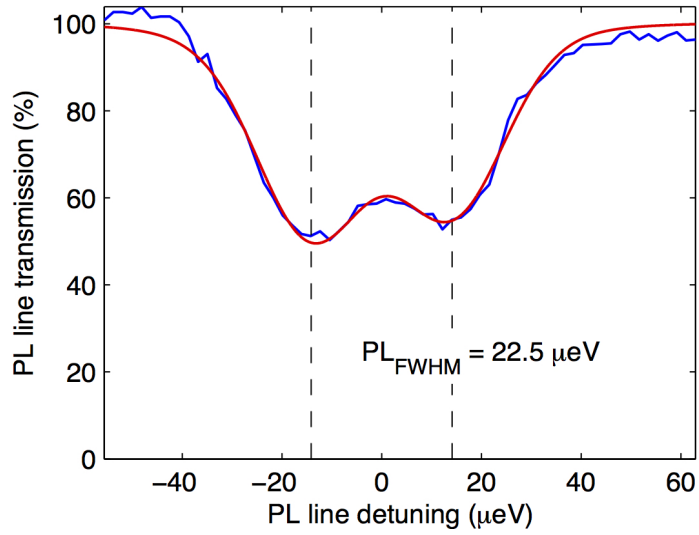

**Fig. S5 | Emission linewidth.** Transmission of the PL line through the rubidium vapour: experimental data (blue) and fit of our model (red). The experimental data is taken from the Scan part of Fig. 3d. The data show two overlapping dips that correspond to the $D_2$ hyperfine structure, indicated by dashed lines. The fit reveals the emission linewidth with full width at half maximum (FWHM) of 22.5 ± 0.6 µeV. In the fit, we model a QD emission line as a Gaussian. Linewidth of the emission, the vapour temperature and the scale of PL line detuning are the fit parameters.

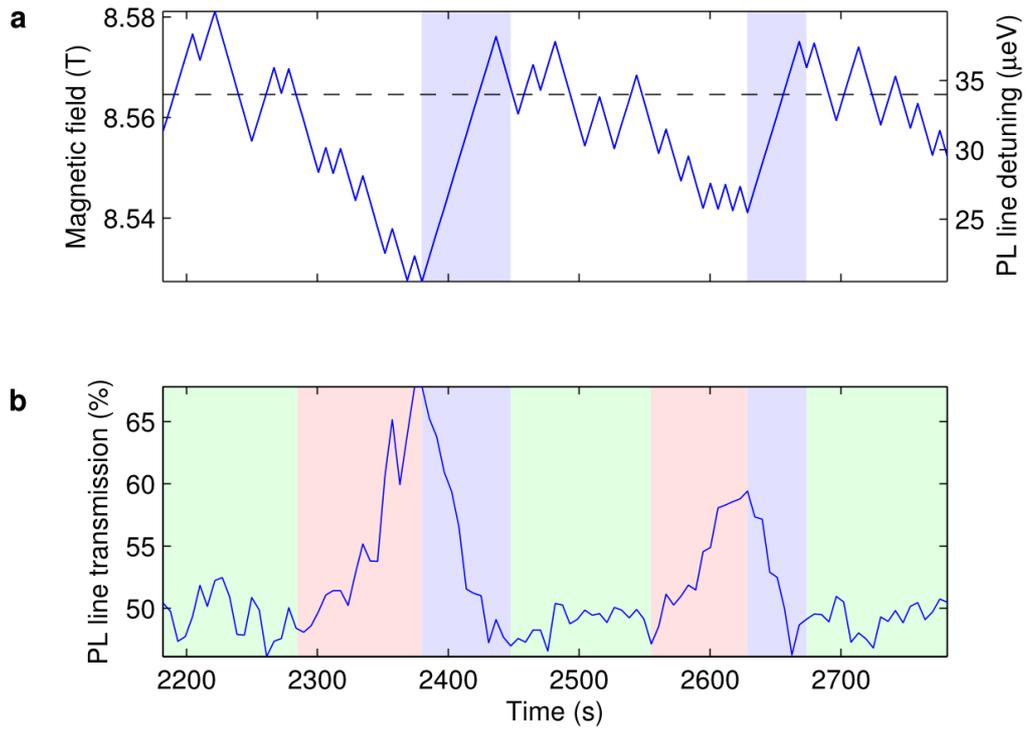

**Fig. S6 | Suppression of spontaneous charge fluctuations.** Zoom-in on the data in Re-lock part of Fig. 3a and 3d. **a,** Modulation of the magnetic field applied to the sample, expressed also as the detuning of the PL line from a rubidium transition, indicated by the dashed line. **b,** Transmission ratio of the normalized signal and the reference lines. The green areas indicate the time intervals when the QD is locked to a rubidium transition. The red areas mark the periods of spontaneous detuning, due to charge fluctuations in the vicinity of the QD. The blue areas indicate the phase when the feedback algorithm successfully counteracts such spontaneous detuning, and re-locks the QD back to the rubidium transition.

# Supplementary References

S1. Atkinson, P., Zallo, E. & Schmidt, O. G. Independent wavelength and density control of uniform GaAs/AlGaAs quantum dots grown by infilling self-assembled nanoholes. *J. Appl. Phys.* **112**, 054303 (2012).